\documentclass{emulateapj}

\slugcomment{ver.Feb.14; draft 1 pages}
\slugcomment{ver.Apr.16; draft 8 pages}
\slugcomment{ver.Jul.18; draft 12 pages}
\slugcomment{ver.Aug.08; draft 13 pages}
\slugcomment{ver.Aug.22; revisions after the 1st circulate}
\slugcomment{ver.Sep.19; almost complete ver.1}
\slugcomment{ver.Sep.26; after textcheck}
\slugcomment{ver.Nov.4; ver.2}
\slugcomment{ver.Nov.15; ver.3}
\slugcomment{Received 2013 September 29 2013; accepted 2013 November 18}

\shorttitle{Extended Ly$\alpha$ emission from a high-$z$ DLA}
\shortauthors{Kashikawa et al.}

\begin{document}


\title{Extended Ly$\alpha$ emission from a damped Ly$\alpha$ absorber at z=3.115\altaffilmark{1}}


\author{
Nobunari Kashikawa\altaffilmark{2,3}, 
Toru Misawa\altaffilmark{4},
Yosuke Minowa\altaffilmark{5},
Katsuya Okoshi\altaffilmark{6},
Takashi Hattori\altaffilmark{5},
Jun Toshikawa\altaffilmark{3},
Shogo Ishikawa\altaffilmark{3},
and
Masafusa Onoue\altaffilmark{3}
}

%
%
%

\email{n.kashikawa@nao.ac.jp}


\altaffiltext{1}{Based on data collected at the Subaru Telescope, which is operated by the National Astronomical Observatory of Japan}
\altaffiltext{2}{Optical and Infrared Astronomy Division, National Astronomical Observatory, Mitaka, Tokyo 181-8588, Japan.}
\altaffiltext{3}{Department of Astronomy, School of Science, Graduate University for Advanced Studies, Mitaka, Tokyo 181-8588, Japan.}
\altaffiltext{4}{School of General Education, Shinshu University, 3-1-1, Asahi, Matsumoto, Nagano 390-8621, Japan}
\altaffiltext{5}{Subaru Telescope, National Astronomical Observatory of Japan, 650 North A'ohoku Place, Hilo, HI 96720.}
\altaffiltext{6}{Tokyo University of Science, Tomino 102-1, Oshamambe, Hokkaido, 049-3514, Japan}

\begin{abstract}

We searched for star formation activity associated with high-$z$ Damped Ly$\alpha$ systems (DLAs) with Subaru telescope.
We used a set of narrow-band (NB) filters whose central wavelengths correspond to the redshifted Ly$\alpha$ emission lines of targeted DLA absorbers at $3<z<4.5$.
We detected one apparent NB-excess object located $3.80$ arcsec ($\sim28h_{70}^{-1}$kpc) away from the quasar SDSS J031036.84+005521.7. 
Follow-up spectroscopy revealed an asymmetric Ly$\alpha$ emission at $z_{em}=3.115\pm0.003$, which perfectly matches the sub-DLA trough at $z_{abs}=3.1150$ with log$N$(H{\sc i})/cm$^{-2}=20.05$.  
The Ly$\alpha$ luminosity is estimated to be L$_{Ly\alpha}=1.07\times10^{42}$ erg s$^{-1}$, which corresponds to a star formation rate of $0.97$ M$_\odot$ yr$^{-1}$.
Interestingly, the detected Ly$\alpha$ emission is spatially extended with a sharp peak.
The large extent of the Ly$\alpha$ emission is remarkably one-sided toward the quasar line-of-sight, and is redshifted.
The observed spatially asymmetric surface brightness profile can be qualitatively explained by a model of a DLA host galaxy, assuming a galactic outflow and a clumpy distribution of H {\sc i} clouds in the circumgalactic medium.
This large Ly$\alpha$ extension, which is similar to those found in \citet{rau08}, could be the result of complicated anisotropic radiative transfer through the surrounding neutral gas embedded in the DLA.

\end{abstract}


\keywords{cosmology: observation --- galaxies: high-redshift --- galaxies: galaxies --- intergalactic medium --- quasars: absorption lines}




\section{Introduction}

Damped Ly$\alpha$ systems (DLAs) are the highest column density [$N$(H {\sc i})$\geq2\times10^{20}$ cm$^{-2}$] absorbers identified in the spectra of quasars, and are known to be major contributors to the neutral hydrogen gas available for vigorous star formation in early epochs.
The majority of neutral hydrogen at $z>3$ resides in DLAs \citep{per05}; therefore tracing their evolution over cosmic time is a complementaly approach towards understanding galaxy formation.
Which galaxy populations harbor DLA systems and how do they evolve with time? 
Based on the similarity of H {\sc i} contents and gas kinematics, DLAs have been suggested as possible progenitors of the present-day spiral galaxies \citep{pro97}; however, the [Zn/H] abundance distribution and its evolution are inconsistent with the hypothesis \citep{pet04}.
Other interpretations of possible DLA candidates include gas-rich dwarf galaxies, merging protogalactic clouds, building blocks of current galaxies, or outflows from protogalaxies. 
Deep imaging for DLAs at $z<1$, where direct detection is easier, reveals diverse morphologies in the galaxy counterparts of DLA absorbers (hereafter ^^ ^^ DLA galaxies"), including dwarf galaxies, low-luminosity galaxies, and normal disk galaxies (e.g., \citealp{che03, che05, rao11}).
Concerning the counterpart of high-$z$ ($z>2$) DLA systems, there are significantly fewer identifications (e.g., \citealp{kro12, fyn13} and references therein.), but also here the detected counterparts display diverse morphologies.
Studying the conversion process of gas into stars in a hierarchically evolving galactic halo can improve our understanding of galaxy formation, and constrain structure formation scenarios.
Therefore, it is quite important to detect emissions from the DLA galaxies, followed by measuring the star formation rate (SFR) of this large gas reservoir.
Although absorption studies of DLAs revealed their gas contents and chemical abundances, observations of DLA emissions would additionally tell us about their SFRs.
However, the SFR estimate in DLAs has proven to be very difficult, because the stellar counterparts of DLAs are often too faint to be detected in emissions, except few cases (e.g., \citealp{mol02, kro13, fyn13}).
DLA galaxies have been identified at $z>3$ in only two cases \citep{djo96, sch12}.

On the other hand, the Lyman break technique has been remarkably successful at finding galaxies in the early universe at $z>3$.
The number of spectroscopically confirmed Lyman-break galaxies (LBGs) has grown to more than $10^3$.
Their bright UV continuum flux suggests active ongoing star formation, and the large LBG sample enables us to constrain the cosmic SFR history.
However, metal enrichment in the gas can be investigated for only a few bright LBGs with the help of gravitational lensing (e.g., \citealp{sma07}).
Both LBGs and DLAs are currently effective tracers of galaxy evolution at high-$z$, but it remains unclear how these populations are related to each other.
The simple interpretations are that DLAs are the early-stage reservoirs of gas from which the stars in LBGs are forming, the sub-$L$* population of LBGs or Lyman $\alpha$ emitters (LAEs)\citep{rau08, yaj12}, or are arising in the outer regions of the LBG galactic halo \citep{wol05}.
The measured emission properties of DLA galaxies at $z\sim3$ fall within the range of emission properties for LBGs of the same luminosity \citep{mol02}; the same holds for $[$O {\sc iii}$]$ properties \citep{wea05}.
High-metallicity DLAs are naturally expected to have luminous galaxy counterparts because of its metallicity-luminosity relation \citep{fyn08, fyn13}
However, the small sample of identified DLA galaxies still critically limits any systematic comparison between DLAs and LBGs/LAEs. 
The DLA-LBG connection is also attracting a great deal of attention in terms of a unique approach to measure the halo mass of DLAs from the DLA-LBG clustering amplitude \citep{coo06}.
Our understanding of the physical or evolutionary connections between these two distinctly selected populations, i.e., H {\sc i} cross section-selected vs. color selected, in the early universe is still in its infancy.

The galactic outflow is another aspect of the connection between DLAs and star-forming galaxies.
Large-scale galactic outflows are ubiquitously found in LBGs and LAEs at $z\sim3$ (e.g., \citealp{sha03, jon12, cho13}). 
These outflows, which are generally referred to as ^^ ^^ feedback" from star-formation activity, are powered by supernovae, stellar winds, photons from young stars, or some combination of the above, leading to self-regulated growth.
Outflow driven by starbursts or active galactic nuclei significantly blows out gas from the galaxy, carrying metal-polluted material to the surrounding IGM. 
This galactic feedback plays an important role in the history of star formation and the early universe.
The galactic wind, which expels gas from galaxies and modifies their chemical evolution by preferentially ejecting metals from those with lower masses, is generally regarded as a plausible origin of the mass-metallicity relation \citep{tre04, led06}.
Surprisingly, the IGM metallicity, which was primarily enriched by galactic winds from galaxies, showed almost no evolution during $2<z<6$ \citep{son04} and a possible downturn at $z>6$ \citep{rya09, sim11}. 
This implies that the IGM at $z>6$ had already been sufficiently metal-enriched by feedback from early galaxies. 
In fact, recent studies (e.g., \citealp{rau08, not12, kro13}) indicate that outflowing gas is associated with the DLA galaxies.
Theoretically, \citet{nag04} suggested that strong galactic winds are required to reproduce the observed H{\sc i} column density distribution function.

As mentioned above, only few high-$z$ DLA galaxies have been detected in emissions so far.
In this study, we carried out a search for star formation activity associated with DLAs with Subaru telescope.
We used a set of narrow-band (NB) filters, whose sensitive wavelengths corresponded to the redshifted Ly$\alpha$ emission lines of target DLA absorbers at $z>3$.
In deep broad-band imaging, it is too difficult to detect faint objects near the line-of-sight (LOS) of bright quasars, and detections are too vague to confirm the redshift of the detected object.
The excellent image quality of HST has been successful at identifying faint candidates close to the quasar sight lines, but redshift determination is quite difficult (e.g., indistinguishable from the quasar host galaxy, or in the case of multiple detections of candidates, see \citealp{war01}).
In long-slit spectroscopy, a blind slit alignment has a slim chance of encountering the target DLA galaxy, and slit scanning would be time-consuming.
The advantages of NB imaging for Ly$\alpha$ emissions of DLA galaxies are that high-sensitivity for emissions can be obtained and the background quasar light around the DLA trough is blocked.
This is essential for the direct detection of DLA galaxies that have small impact parameters. 
Most of the previous detections of high-z DLA galaxies have been found as LAEs, which can be easily understood given that the DLA galaxy are young, gas-rich, less dusty, and undergoing an initial star burst.
The DLA galaxies could be faint in continuum flux and fairly bright in emissions, which can be efficiently detected using NB imaging.
There are previous NB searches for DLA galaxy \citep{fyn03,gro09}, though they failed to detect a counterpart galaxy.


This paper is organized as follows:
In \S~2, we describe the NB imaging observation to search for DLA galaxy candidates.
In \S~3, we describe results of the search for the NB-excess objects, in which we found one good candidate.
In \S~4, we present follow-up spectroscopy to confirm that the candidate is actually associated with the target DLA. 
We found an extended Ly$\alpha$ emission, the possible interpretations of which are discussed in \S~5.

Throughout the paper, we assume a flat $\Lambda$CDM cosmology with $\Omega_{\rm m}=0.27$, $\Omega_\Lambda=0.73$, and $H_0=70$ $h_{70}$ km s$^{-1}$ Mpc$^{-1}$. 
These parameters are consistent with recent CMB constraints \citep{kom11}.
Magnitudes are given in the AB system and distances are quoted in proper units, unless otherwise denoted.

\section{Imaging observation and photometry}

We observed $5$ DLA fields with Subaru/FOCAS \citep{kas04}.
The target DLAs were selected from the DLA catalog based on SDSS DR3 \citep{pro05} by the following criteria: 1) $3<z_{\rm abs}<4.5$, 2) the redshifted Ly$\alpha$ emission lines of the target galaxy were within the sensitive wavelength range of any of the FOCAS NB filters, 
3) $\Delta v(z_{\rm abs}-z_{\rm em})>3000$ km s$^{-1}$ to exclude $z_{\rm abs}\sim z_{\rm em}$ DLA systems (e.g., \citealp{mol93}),  that are likely to be different from intervening DLAs due to their special environments \citep{mol98, ell10}.
The FOCAS NB filters were carefully designed to avoid the strong OH night sky emission lines.
The targets are listed in Table \ref{tab_dla}.
Q2233+131, in which a DLA galaxy at $z=3.15$ \citep{djo96} was already detected, was also included in the target list to see the feasibility of our approach.

\begin{deluxetable*}{rlrlrrrr}
\tablecaption{Summary of targets \label{tab_dla}}
\tablewidth{0pt}
\tablehead{
\colhead{Quasar} & \colhead{$z_{abs}$} &  \colhead{$z_{em}$} & \colhead{log$N$(H{\sc i})} & \colhead{$Q_{rmag}$} & \colhead{NB} & \colhead{$\lambda_c$(\AA)} & \colhead{$\delta\lambda$(\AA)}
}
\startdata      
Q2233+131           & 3.151 & 3.296 & 19.95 & 18.37 & NB502 & 5025 & 60 \\
SDSS J031036.84+005521.7 & 3.114$^a$ & 3.782 & 20.20$^b$ & 19.71 & NB502 & 5025 & 60 \\
SDSS J001240.57+135236.7 & 3.022 & 3.187 & 20.55 & 19.61 & NB487 & 4882 & 53 \\
SDSS J162626.50+275132.4 & 4.495 & 5.275 & 21.35 & 21.65 & NB670 & 6681 & 85 \\
SDSS J224147.76+135202.7 & 4.283 & 4.448 & 21.15 & 19.50 & NB642 & 6428 & 127 
\enddata
\tablenotetext{a}{This value is taken from SDSS DR5. This study evaluated as $3.1150\pm0.0001$.}
\tablenotetext{a}{This value is taken from SDSS DR5. This study evaluated as $20.05\pm0.05$.}
\end{deluxetable*}

FOCAS has two $2$k $\times 4$k MIT/LL CCDs, and covers a circular area with a diameter of $6'$ with a pixel scale of $0.''104$ pixel$^{-1}$.
Observations were made on the nights of UT $2005$ August 26 and September 23-24.
The total integration time for the NB filters was $2400$-$7200$ s and the typical unit exposure time was $1200$ s.
We also took short exposures with broadband filters of Johnson $B$, $V$, and $R$ bands, to constrain the continuum flux of DLA galaxies, and to discriminate LAEs from low-$z$ H$\alpha$, $[$O {\sc iii}$]$ and $[$O {\sc ii}$]$ emitters. 
The two BB filters for each target were chosen to trace the blue (BB$_b$) and red (BB$_r$) side wavelengths of the NB.   
BB imaging also helped to catch the strong Lyman break in the spectral energy distributions, allowing us to discriminate from the low-$z$ emission line galaxies.
The typical unit exposure time was $900$ s for $B$ and $240$ s for $V$ and $R$; shorter exposures were used for redder filters because of the increase in sky brightness with wavelength.
The sky conditions were fairly good with a seeing size of $0.4$-$0.8$ arcsec.
The journal of observations is provided in Table \ref{tab_img}.

\begin{deluxetable*}{rrrrrrr}
\tablecaption{Jounal of observations \label{tab_img}}
\tablewidth{0pt}
\tablehead{
\colhead{Quasar} & \colhead{date(UT)} &  \colhead{NB} & \colhead{$T_{\rm integ}^{\rm NB}$(s)} & \colhead{BB} & \colhead{$T_{\rm integ}^{\rm BB}$(s)} & \colhead{seeing(arcsec)}  
}
\startdata      
Q2233+131           & 26/08/05 & NB502 & 6000 & B, V & 180, 120 & 0.5   \\
SDSS J031036.84+005521.7 & 24/09/05 & NB502 & 7200 & B, V & 900, 600 & 0.4   \\
SDSS J001240.57+135236.7 & 23/09/05 & NB487 & 2400 & B, V & 900, 600 & 0.8   \\
SDSS J162626.50+275132.4 & 26/08/05 & NB670 & 6000 & V, R & 900, 600 & 0.6   \\
SDSS J224147.76+135202.7 & 26/08/05 & NB642 & 4800 & V, R & 600, 600 & 0.5   
\enddata
\end{deluxetable*}

The data were reduced in the standard manner following the FOCAS data reduction pipeline.
Photometric calibration was made with photometric standard stars, PG1657+078 and PG0231+051 for the BB bands, and spectroscopic standard stars PG0205+134, PG1545+035, and PG1708+602, for the NB bands.
We performed object detection and photometry by running SExtractor version 2.8.6 \citep{ber96} on the images.
Object detection was made in the NB band images.
For all objects detected in a given band pass, the magnitudes and several other parameters were measured in the other band passes at exactly the same positions as in the detection-band image, using the {\lq}double image mode{\rq} of SExtractor.
We detected objects that had $6$ connected pixels above $2\sigma$ of the sky background rms noise and took photometric measurements at the $2\sigma$ level.
Aperture photometry was performed with a $2\arcsec \phi$ aperture to derive the colors of the detected objects.

Since we were looking for faint objects near bright target quasars, object detection and photometry were inefficient and unreliable.
We applied GALFIT \citep{pen10} to detect the faint objects after removing the bright quasar image.
A two-dimensional Moffat profile was fit to the quasar images.
We searched for NB-excess (NBe) objects around the target quasar LOS within 20\arcsec radius.
The NBe criterion due to Ly$\alpha$ emission was defined as

\begin{equation}
{\rm NBe}\tbond {\rm BB}_{\rm cont}-{\rm NB}>1.0,
\end{equation}

where $BB_{\rm cont}$ is the continuum magnitude at the NB central wavelength. 
It can be roughly estimated from

\begin{equation}
{\rm BB}_{\rm cont}=\frac{{\rm BB}_b\times(\lambda^{BB_r}_c-\lambda^{NB}_c)+ {\rm BB}_r\times(\lambda^{NB}_c-\lambda^{BB_b}_c)}{\lambda^{BB_r}_c-\lambda^{BB_b}_c},
\end{equation}

where $\lambda^{BB_r}_c$, $\lambda^{NB}_c$, and $\lambda^{BB_b}_c$ are the central wavelengths of the BB$_r$, NB, and BB$_b$ bands, respectively.

\section{DLA galaxy candidates}

Out of $5$ targets, we detected one apparent NB-excess object near to SDSS J031036.84+005521.7.
As expected, we also detected the NB-excess object from Q2233+131, in which a DLA galaxy has already been detected by \citet{djo96}.
We also detected a very red object very near to the quasar SDSS J001240.57+135236.7, though the quick follow-up spectroscopy revealed that it was a nearby overlapping object.
We did not detect any NB-excess objects around two other targets within 20\arcsec from the quasar-LOS. 
Notes on three individual detections follow.

\begin{figure}
\vspace*{-0.1cm}
\epsscale{1.2}
\plotone{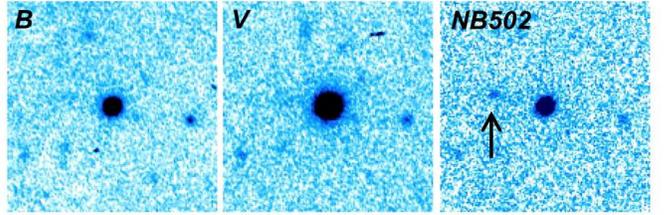}
\vspace*{-2.8cm}
\epsscale{1.0}
\caption{The images of 15\arcsec $\times$ 15\arcsec surrounding the centered quasar SDSS J031036.84+005521.7 in B, V, and NB502 (from left to right). N is up and E is to the left. The DLA galaxy is seen east-northeast of the quasar at an impact parameter of 3.\arcsec8. 
\label{fig_j0310}}
\vspace*{0.5cm}
\end{figure}

\subsection{SDSS J031036.84+005521.7, $z_{abs}=3.114$}

This sub-DLA was listed at $z_{abs}=3.1142$ with log$N$(H{\sc i})$=20.20\pm0.15$ in SDSS DR5.
We detected an NB-excess object, which did not appear in the $B$ filter (the $3\sigma$ limiting magnitude in $2\arcsec$ aperture $<25.32$) and $V$ ($<25.50$) images, $3.80$ arcsec away from the quasar LOS in east-northeast direction as illustrated in Figure \ref{fig_j0310}.
The NB total magnitude is NB$502=25.46\pm0.13$, and the NB excess is NBe$>1.54$ ($1\sigma$).
We found no additional NB-excess objects around 20\arcsec from the quasar LOS.
The NB-excess object was discovered at a relatively large impact parameter of $\sim28h_{70}^{-1}$kpc away from the DLA\footnote{This absorber should strictly be classified as a sub-DLA population based on its definition, [$10^{19}< N$(H {\sc i})$<2\times10^{20}$ cm$^{-2}$]
; however for simplicity, we will call it hereafter a ^^ ^^ DLA''.}.
The NB emission is sufficiently bright to carry out follow-up spectroscopy, as discussed later.

\begin{figure}
\vspace*{-0.1cm}
\epsscale{1.2}
\plotone{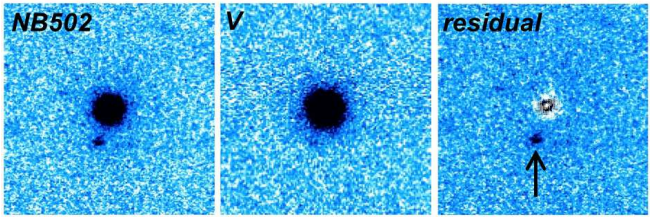}
\vspace*{-2.8cm}
\caption{The images of 15\arcsec $\times$ 15\arcsec surrounding the centered quasar Q2233+131 in NB502, V, and the residual NB502-band image after subtracting the quasar model using GALFIT (from left to right). N is up and E is to the left. The DLA galaxy is seen south-east of the quasar at an impact parameter of 2.\arcsec4.    
\label{fig_q2233}}
\vspace*{0.5cm}
\end{figure}

\subsection{Q2233+131, $z_{abs}=3.151$}

The DLA was listed at $z_{abs}=3.150$ with log$N$(H{\sc i})$=20.2$ in \citet{cur02}.
It was identified on J223619.19+132620.3 at $z_{abs}=3.151$ with log$N$(H{\sc i})$=20.0$ in SDSS DR5.
The metallicity was estimated as [Fe/H]=-1.4 \citep{lu97}.
The galaxy was first detected by \citet{ste95}, and it was confirmed as a galaxy counterpart of the DLA by \citet{djo96}.
As shown in Figure \ref{fig_q2233}, we detected an NB-excess object $2.41$ arcsec south-east of the quasar LOS, in agreement with \citet{djo96}.
The NB total magnitude was NB$502=24.30\pm0.04$, and NBe$=1.27$.

\citet{chr04} detected Ly$\alpha$ emission from the DLA galaxy at Q2233+131 using a Integral Field Spectrograph, and derived its luminosity, SFR, and size.
It is notable that the Ly$\alpha$ emission nebula they found is significantly extended, with a size of $23\times38$ kpc, which has an appearance similar to the Ly$\alpha$ blobs of LBGs, 
but is more diffuse and less extended.
As they suggested, it was expected that the detectable size of the extended Ly$\alpha$ nebula would be larger when observed in fainter sensitivity.
However, we could not detect such an extension in our Ly$\alpha$ image.
The $3\sigma$ limiting surface brightness of our NB image is $3.5\times10^{-17}$ erg s$^{-1}$ cm$^{-2}$ arcsec$^{-2}$, which is comparable to their sensitivity of $4\times10^{-17}$ erg s$^{-1}$ cm$^{-2}$ arcsec$^{-2}$.  
The excellent image quality of Subaru revealed that the DLA galaxy has a rather compact Ly$\alpha$ structure. 
Based on GALFIT analysis, the Ly$\alpha$ image of the DLA galaxy has an effective radius of $R_e=1.35$ arcsec $=10.26$ kpc with a Sersic index of $n=1.80$. 
The reason for the discrepancy between our observation and \citet{chr04} is not clear at this moment.

\begin{figure}
\vspace*{-0.1cm}
\epsscale{1.2}
\plotone{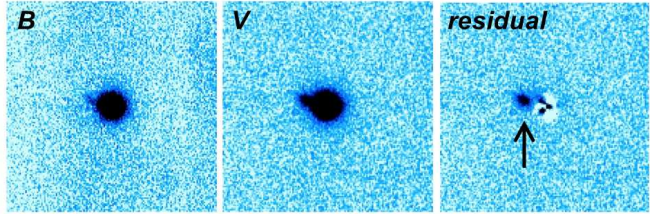}
\vspace*{-2.8cm}
\epsscale{1.0}
\caption{The images of 15\arcsec $\times$ 15\arcsec surrounding the centered quasar SDSS J001240.57+135236.7 in B, V, and the residual V-band image after subtracting the quasar model using GALFIT (from left to right). N is up and E is to the left. The DLA galaxy candidate is seen east of the quasar at an impact parameter of 1.\arcsec5. 
\label{fig_j0012}}
\vspace*{0.5cm}
\end{figure}

\begin{figure}
\vspace*{-1.4cm}
\epsscale{1.2}
\plotone{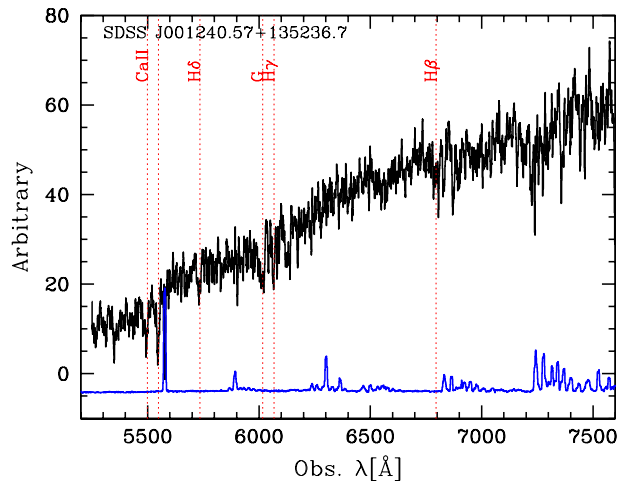}
\vspace*{-1.3cm}
\epsscale{1.0}
\caption{The spectrum of the DLA galaxy candidate found in SDSS J001240.57+135236.7 image. It was found to be a red galaxy at $z=0.398$, based on the CaII, H$\delta$, G, H$\gamma$, and H$\beta$ absorption lines. 
\label{fig_j0012sp}}
\end{figure}

\subsection{SDSS J001240.57+135236.7, $z_{abs}=3.022$}

No NB-excess objects are found around the quasar LOS; however, there is an apparent overlapping object on the quasar (Figure \ref{fig_j0012}).
This was an obvious DLA galaxy candidate quickly found during our observation; therefore, we switched the observing mode to spectroscopy to verify its spectrum.
The FOCAS observation was first made with the VPH520 grating, which covers $4450$\AA$--6050$\AA~; however, we did not obtain any signal except at long wavelengths.
Then, we obtained a spectrum with the VPH650 grating and the Y47 order-cut filter.
The spectra covered $5300-7700$ \AA, with a pixel resolution of $0.63$ \AA.
The $0\arcsec.8$-wide slit gave a spectroscopic resolution of $R\sim1300$.
The integration time was 1 hr.
We did not take flux calibration data for spectrophotometric standard stars.
The resultant spectrum is shown in Figure \ref{fig_j0012sp}.
This object was found to be a red galaxy at $z=0.398$, based on our identification of CaII, H$\delta$, G, H$\gamma$, and H$\beta$ absorption lines. 
We concluded that this low-$z$ galaxy happened to overlap to the quasar LOS, and is not responsible for the DLA at  $z_{abs}=3.022$.      

\begin{figure}
\epsscale{1.3}
\plotone{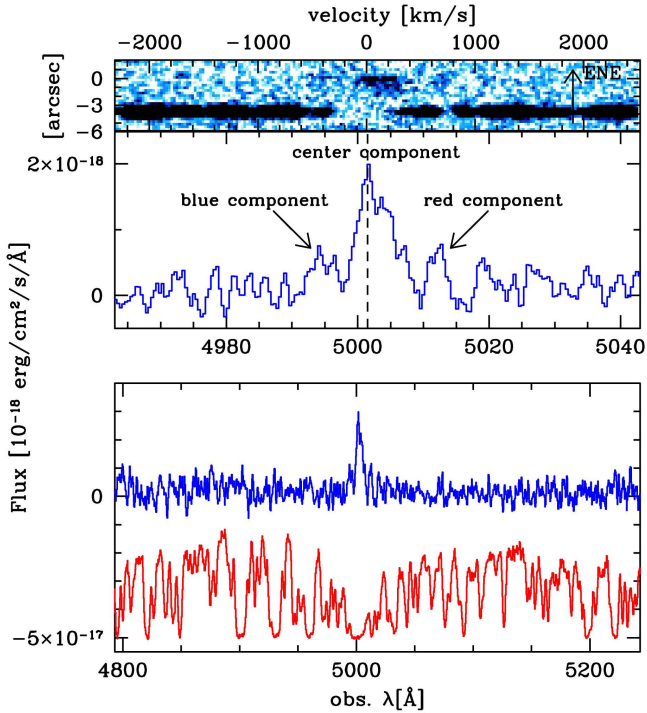}
\caption{The upper panel shows the 1D and 2D spectrum around the Ly$\alpha$ emission of the galaxy corresponding the DLA at $z_{abs}=3.115$ of the quasar SDSS J031036.84+005521.7. 
The thick line is the quasar spectrum in the 2D spectrum. The asymmetric Ly$\alpha$ line is clearly visible at $\sim5000$\AA~, offset from the quasar by 3.8 arcsec.
Vertical direction is east-northeast (ENE). 
The lower panel shows the comparison of the spectra of the DLA galaxy (upper, blue line) to the quasar (lower, red line). 
The flux of galaxy spectrum was magnified by $\times15$, and the quasar spectrum is shifted downward by $-1\times10^{-17}$ erg s$^{-1}$ cm$^{-2}$ \AA$^{-1}$ for clarification.
\label{fig_j0310sp}}
\end{figure}

\begin{figure*}
\epsscale{1.2}
\plotone{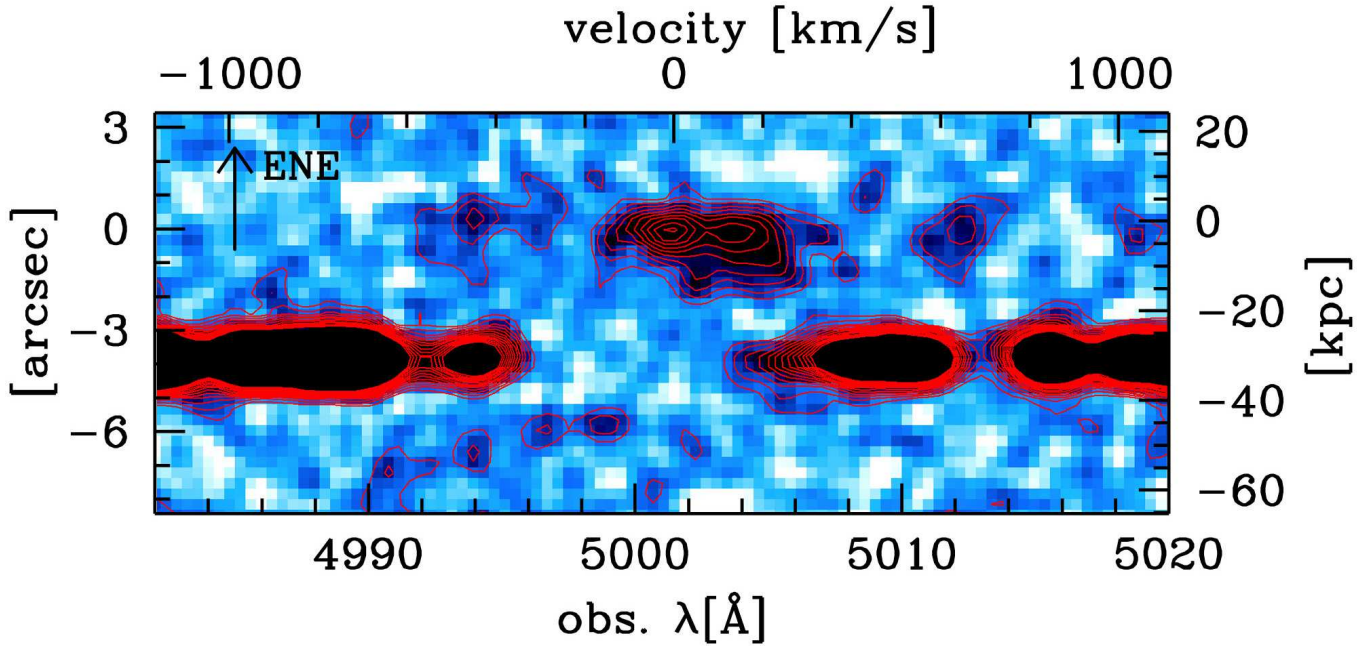}
\vspace*{-9.3cm}
\epsscale{1.0}
\caption{The close-up 2D spectrum around the Ly$\alpha$ emission at $5000$\AA~of the galaxy corresponding to the DLA at $z_{abs}=3.115$ of the quasar SDSS J031036.84+005521.7.
Vertical direction is east-northeast (ENE). 
The spectrum has been smoothed with a $3\times3$ pixel top hat filter. 
The red contours show flux density with a width of $2.5\times10^{-19}$ erg s$^{-1}$ cm$^{-2}$ \AA$^{-1}$. 
The outermost contour corresponds to a flux density of $2.5\times10^{-19}$ erg s$^{-1}$ cm$^{-2}$ \AA$^{-1}$, which corresponds to a $1.5\sigma$ background fluctuation. 
\label{fig_j0310cont}}
\vspace*{0.5cm}
\end{figure*}

\section{Spectroscopy}

The spectroscopic data of the DLA galaxy candidate at SDSS J031036.84+005521.7 were obtained on UT 2007 October 3 using FOCAS at the Subaru telescope.  
The observation was made with the VPH520 grating, which covers $4450$\AA$-6050$\AA~ with a pixel resolution of $0.39$\AA.
We placed the quasar and the DLA galaxy candidate simultaneously along the long-slit.
The $0\arcsec.8$-wide slit, which gave a spectroscopic resolution of $R\sim1500$, was placed so as to penetrate both the NB-excess object and the quasar.
The spatial resolution was $0\arcsec.3$ pixel$^{-1}$ with $3$-pixel on-chip binning.
Dithering of $1.\arcsec0$ was performed during the observation to achieve good background subtraction.
The seeing size was from 0.\arcsec5 to 1.\arcsec0 and the integration time was 3.5 hrs.
The data were reduced using standard techniques following the FOCAS data reduction pipeline.
The final spectrum was constructed from the median frame, and the flux calibration was made with spectroscopic standard star Feige 110.

Figure \ref{fig_j0310sp} shows the spectrum around the DLA trough at $z_{abs}=3.114$ of the quasar SDSS J031036.84+005521.7.
The 2D spectrum (upper panel) shows a clear emission line signal at $\sim5000$\AA~, offset from the quasar (thick line) by $3.8$ arcsec, which corresponds to the spatial position of the NB-excess DLA galaxy candidate.
The object shows an apparent asymmetric emission line profile, which is characteristic of high-z Ly$\alpha$ emission.
The weighted skewness parameter, $S_W$, which is a quantitative estimate the line asymmetry \citep{kas06}, was estimated to be $S_W=3.103 \pm 0.213$, 
suggesting that this is certainly Ly$\alpha$ emission at $z_{em}=3.115\pm0.003$, which is determined from the peak wavelength of the Ly$\alpha$ emission after converted to a vacuum scale and corrected to the heliocentric rest frame, as we did in the Voight fitting to the DLA trough shown below. 
The Ly$\alpha$ luminosity is estimated to be L$_{Ly\alpha}=1.07\pm0.14\times10^{42}$ erg s$^{-1}$, which corresponds to a SFR of $0.97\pm0.13$ M$_\odot$ yr$^{-1}$ \citep{ken98}.
Interestingly, the asymmetric Ly$\alpha$ emission of the galaxy shows an apparent extended structure toward the quasar LOS on the 2D spectrum and a small bump at the blue and red side (Figure \ref{fig_j0310cont}), which implies that a galactic wind might be associated with the galaxy.
We will discuss this possibility in the next section.

We also derived line parameters by fitting a Voigt profile to the DLA trough in the 1D spectrum shown in Figure \ref{fig_j0310sp} (lower panel). 
The spectrum was normalized by dividing out the quasar continuum, which we fit directly with a third-order cubic spline function.  
As the DLA is heavily blended with the Ly$\alpha$ forest, we referred only to the local flux peak and then interpolated between them while fitting.  
No other lines in the Lyman series were covered in the spectrum.
We fitted only the Ly$\alpha$ trough with a single Voigt profile. 
Our best fit parameters were $\log N_{HI}$ = 20.05$\pm$0.05 and $z$ =3.1150$\pm$0.0001.  
The errors are only lower limits because these values do not include the systematic error related to the continuum
fit uncertainty. 
Our results are almost consistent with those reported in the SDSS DLA Survey ($\log N_{HI} = 20.20\pm0.15$ and $z$ = 3.1142)\footnote{http://www.ucolick.org/$\sim$xavier/SDSSDLA}.

\section{Results and Discussion}

\subsection{Galaxy counterpart of DLA at $z_{abs}=3.115$}

Given the almost perfect coincidence of redshifts, we concluded that the Ly$\alpha$ emission comes from the galaxy associated with the DLA.
The velocity difference between the DLA and the Ly$\alpha$ emission is within the spectroscopic resolution of $220$ km s$^{-1}$.
Unfortunately, our relatively low resolution spectrum did not resolve any low-ionization absorption lines to precisely determine the DLA redshift.
It should also be noted that the redshift estimate of Ly$\alpha$ emission may be affected by the H {\sc i} gas absorption in the galaxy. 
The H {\sc i} column density (log$N$(H{\sc i})$=20.05$) at large impact parameter of $\sim28h_{70}^{-1}$kpc is consistent with an H {\sc i} density profile of low-$z$ DLAs \citep{che03, rao11}.
\citet{bou13} detected a DLA galaxy at z=2.33 with comparably large ($\sim26$kpc) impact parameter.
It is also consistent with the DLA model prediction that DLAs with lower gas density are expected to be discovered larger impact parameter (e.g., \citealp{yaj12}).
Recent studies have detected quite a number of DLA galaxies at $z>2$, intentionally targeting for metallicity-rich DLAs (e.g., \citealp{fyn13}).
It is interesting to compare the metallicity between our detection and these studies; however, there has been no metallicity measurement for the DLA at $z_{abs}=3.115$ of SDSS J031036.84+005521.7.

\subsection{Origin of extended Ly$\alpha$ emission}

\begin{figure}
\epsscale{1.2}
\plotone{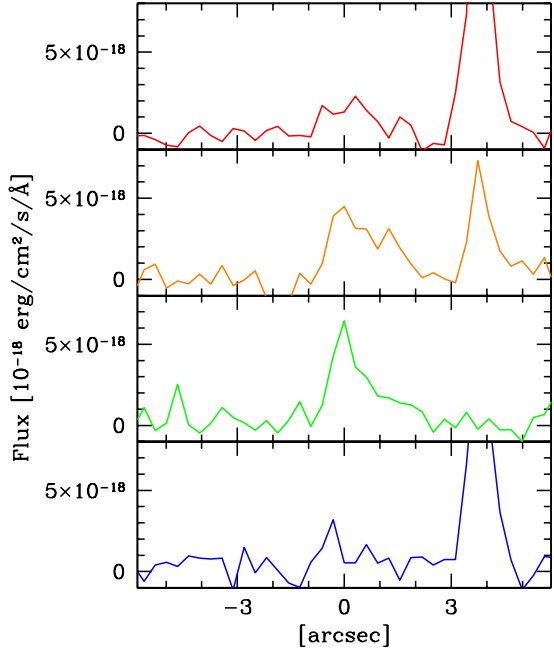}
\caption{Spatial profile of Ly$\alpha$ emission of the DLA galaxy in SDSS J031036.84+005521.7. The four profiles in different wavelength ranges are shown; from bottom to top, $4994\pm2$ (blue component), $5001\pm2$, $5005\pm2$, $5012\pm2$\AA (red component). The quasar spatial profile is seen at +3.8 arcsec position.   
\label{fig_j0310prof}}
\end{figure}

As shown in Figure \ref{fig_j0310cont}, the Ly$\alpha$ emission detected from the DLA galaxy at SDSS J031036.84+005521.7 is spatially extended.
The large extent of the line emission has a sharp peak, whose wavelength perfectly agree with the DLA absorption centroid on the quasar spectrum.  
The spatial extension is remarkably one-sided toward the quasar LOS, and is redshifted.
The 1D-spectrum (Figure \ref{fig_j0310sp}) shows Ly$\alpha$ emission at $5001.65$\AA~with FWHM$=459$ km s$^{-1}$, which we call hereafter ^^ ^^ center'' component.
The maximum spatial extent of the center component is almost 2\arcsec ($\sim$ 15.5 proper kpc ) down to a flux density of $2.5\times10^{-19}$ erg s$^{-1}$ cm$^{-2}$ \AA$^{-1}$, which corresponds to $1.03\times10^{-18}$ erg s$^{-1}$ cm$^{-2}$ \AA$^{-1}$ arcsec$^{-2}$.
The extension is too diffuse, with a surface brightness of $1-5\times10^{-18}$ erg s$^{-1}$ cm$^{-2}$ \AA$^{-1}$ arcsec$^{-2}$, to be identified in the relatively short-exposure NB image.
In addition, there seem to be two possible subcomponents: one at $5012.46$\AA~with FWHM$=188$ km s$^{-1}$ (red component) and the other at $4993.99$\AA~with FWHM$=306$ km s$^{-1}$ (blue component).
Figure \ref{fig_j0310prof} shows the spatial profiles of the extended Ly$\alpha$ emission in four different wavelength ranges. 
The center component (second and third panels from the top) shows an apparently spatially extended profile toward the quasar LOS. 
The red component also seems to have 1.\arcsec2 extension toward the quasar LOS, while the blue component has no apparent spatial extension and its peak position is slightly shifted to the opposite side of the quasar LOS.

Although it is difficult to identify the origin of such extended Ly$\alpha$ emission, we suggest a simple scenario in which this DLA galaxy is accompanied by a galactic wind. 
Given a galactic outflow around the source, the far side of the outflowing cloud could backscatter redshifting Ly$\alpha$ photons, while the blue peak would be strongly absorbed by outflowing neutral hydrogen at the near side. 
Thus, the outflow model generally predicts a double-peaked Ly$\alpha$ line profile in which the red peak is stronger than the blue peak.  
This picture has been well studied using simple spherically symmetric expanding shell models \citep{ver06, ver08}. 
The predicted Ly$\alpha$ profile reproduces the observed one-dimensional spectra of high-z galaxies \citep{kas06, tap07, yam12}.
However, the shell models generally predict a very flat surface brightness profile \citep{bar10}, which appears to be different from that observed showing a sharp peak in this study (Figure \ref{fig_j0310prof}).

Alternatively, the Ly$\alpha$ radiative transfer calculation by  \citep{bar10}, 
which assumes a DLA galaxy with an expanding optically thick gaseous halo, reveals significantly more centrally peaked spatial distribution, and more extended Ly$\alpha$ emission as higher outflow velocities.
According to their model, the central H{\sc i} column density in massive halos is too high to escape the Ly$\alpha$ photons.
The Ly$\alpha$ photon scatter, which is induced by a number of isolated clumps, is enhanced by large bulk motion further from the center.  
Interestingly, their predicted two-dimensional spectrum shows a spatial asymmetry due to the underlying clumpy distribution of H {\sc i} clouds, even assuming a simple spherically symmetric gas distribution and bulk motion.
The observed asymmetric two-dimensional spectrum qualitatively favors this picture.  
Blueshifted and extended Ly$\alpha$ emission is expected when assuming bipolar outflowing gaseous jets perpendicular to the disk; however, the blue Ly$\alpha$ extension cannot be seen on the opposite side (east-northeast direction) of the quasar LOS.
This blueshifted Ly$\alpha$ component could be significantly absorbed by a thick H {\sc i} cloud associated with the galaxy.  
In this case, a coherent, smooth H {\sc i} cloud structure extended over the DLA galaxy and quasar LOS seems unlikely, because a relatively thick DLA trough with log$N$(H{\sc i})$=20.05$ is located across the DLA galaxy from the eastern obscured region.
The eastern obscured part of the galaxy may happen to have a high spatial covering fraction of H {\sc i}.
This is also consistent with the model of clumpy H {\sc i} distribution in a DLA galaxy.
The center and blue components found in Figure \ref{fig_j0310cont}, which have a low signal-to-noise ratio, may correspond to the red and blue peaks in the outflow model, respectively. 
In the case, the rest frame of the DLA galaxy may correspond to the valley between the central and blue components, and is slightly closer to us by $\sim-250$ km s$^{-1}$ than $z=3.115$.
The red component, if real, might be caused by an intermittent process of galactic wind.
Interestingly, a relatively thick absorption corresponding to almost the same redshift as the red component was found in the quasar spectrum.
Such a triple-peaked Ly$\alpha$ emission has been also discovered in \citet{cho13}.
These interpretations reasonably account for the observed extended Ly$\alpha$ emission from the DLA galaxy, though they do not necessarily provide a unique explanation.
Three-dimensional radiative transfer calculations based on an assumed H {\sc i} spatial distribution and the velocity structure of the DLA halo are required for further verification, though it is far beyond the scope of this paper.  

\citet{rau08} found spatially extended Ly$\alpha$ emissions with radii up to $\sim30$ kpc and mean radii of $\sim8$ kpc from the objects, based on a long-exposure (92 hrs) low surface brightness Ly$\alpha$ emitter survey. 
This extension is much larger than the average effective radius of $\sim0.7-3.0$ kpc that \citet{law12a} measured in the optical wavelengths for $z\sim3$ LBGs. 
Such a ubiquitously larger Ly$\alpha$ size compared to the UV continuum has been well obseved at high-z (e.g., \citealp{ste11}) and low-z (e.g., \citealp{hay13}).  
\citet{ste11} found a diffuse extended Ly$\alpha$ emission halo out to $\sim80$ kpc, based on extremely deep stacks of Ly$\alpha$ images.
\citet{rau08} argued that their sample should overlap with the elusive host population of DLAs. 
Although some of their extended Ly$\alpha$ sample shows the signature of a DLA trough on the faint continuum spectrum of the LAE itself \citep{rau11a}, our sample provides the direct evidence that extended Ly$\alpha$ emission actually accompanies the DLA appearing on the nearby quasar LOS.
The DLA galaxy of this study has the Ly$\alpha$ luminosity of $1.07\times10^{42}$ erg s$^{-1}$, which is much fainter than most of other DLA galaxies at $z>2$, but comparable to the bright end of the sample of \citet{rau08}.
The DLA simulation by \citet{bar10} well reproduced spatially extended (up to 50 kpc or larger) faint Ly$\alpha$ emission similar to our observations.
The authors suggested that DLA are predominantly hosted by dark matter halos with $10^{9.5}$-$10^{12}$ M$_\odot$, which is much smaller than those of $L^{*}$-class LBGs.

\subsection{Other interpretations of extended Ly$\alpha$ emission}

\citet{rau11a} found a galaxy with spatially asymmetric Ly$\alpha$ diffuse emission.
Interestingly, their object has blueshifted diffuse emission, which is in clear contrast to our findings.  
They interpreted that the spatially asymmetric emission was caused by partial covering by a DLA cloud in front of the galaxy, and that the blueshift diffuse emission was fluoresced by infalling neutral hydrogen gas from the backside of the DLA host galaxy.    
We did not detect any stellar continuum counterparts of the DLA galaxy in this study, which also invokes a Ly$\alpha$ fluorescence (e.g., \citealp{ade06, can12}); however, no bright nearby ionizing sources have been found.
As discussed in \citet{rau08}, the expected fluorescent Ly$\alpha$ emission signal due to general UV background is extremely low $\sim10^{-19}$erg s$^{-1}$ cm$^{-2}$ arcsec$^{-2}$, which is much lower than our data.
Typical fluorescent surface brightnesses of $>10^{-18}$erg s$^{-1}$ cm$^{-2}$ arcsec$^{-2}$ can be attained only if illuminated by nearby bright quasars \citep{can12}, which is not likely in this case. 
Unlike \citet{rau11a}, the object in this study has neither a disturbed morphology nor any sign of interactions on the NB image. 

Another possible mechanism for extended Ly$\alpha$ emission is due to multiple scattering in the accreting cool gas onto the dark matter halo of the galaxy \citep{hai00, yan06, lat11}.
Following the same prescription in \citet{rau11a}, using an analytic relation between the star-formation produced Ly$\alpha$ luminosity and the halo mass of \citet{fau10}, the observed Ly$\alpha$ luminosity of L$_{Ly\alpha}=1.07\times10^{42}$ erg s$^{-1}$ yields a halo mass of $2.2\times10^{10}$M$_\odot$, comparable to that of \citet{rau11b}.
The expected cooling Ly$\alpha$ radiation from cold accretion is only $4\%$ of stellar Ly$\alpha$ emission at this halo mass, and this cooling is only effective in massive halos.
Therefore, the observed extended Ly$\alpha$ feature is unlikely to be produced by cooling radiation, though the derived halo mass based on Ly$\alpha$ emission is largely affected by kinematics and H {\sc i} distribution in the halo.

Several studies suggested a more direct connection between DLA and galactic wind.
The DLA arises in the LOS where a collimated wind intersects the DLA \citep{nul98}; however this is not the case for this object, because possible open up direction of the wind cone does not correspond to the relative position of the DLA on the LOS. 

\subsection{Outflow and DLA}

Although \citet{wol06} constrains the spatial extent of star formation in DLAs to be less than $<3$ kpc, the large spatial extent observed here is the result of complicated anisotropic radiative transfer through the surrounding neutral gas embedded in the DLA \citep{rau08, bar11}.   
\citet{kro13} also suggested the existence of a galactic wind associated with the DLA galaxy at $z=2.35$.
\citet{not12} proposed a similar picture, in which the DLA galaxy at $z=2.2$ is associated with outflowing gas.
They detected double-peaked Ly$\alpha$ emission along with [O {\sc iii}] and H$\alpha$ emission, providing stronger constraints on the Ly$\alpha$ radiative transfer on the outflow model.
They also argued for possible extended (FWHM$\sim8$kpc) Ly$\alpha$ emission, though the small ($0.9$kpc) impact parameter makes it difficult to reveal the real extent upon the DLA trough.
One of the general difficulties of identifying a DLA galaxy is its small impact parameter.
There are some claims that DLA galaxies could have much smaller impact parameters ($<1$ arcsec) than ever searched \citep{oko05}, which is supported by the idea that the inner region of DLA galaxies would have the same high H {\sc i} column density as the DLA criterion of $N$(H {\sc i})$\geq2\times10^{20}$ cm$^{-2}$.
Our object, at a relatively large separation from the quasar-LOS, may be a rare case that fortuitously revealed such a extended feature. 
It is unclear whether such a low-SFR galaxy can sustain an outflow. 
More accurate SFR measurements based on the H$\alpha$ line, the UV continuum luminosity, or measurements of the virial mass based on object size and velocity dispersion are necessary for further discussion.


\smallskip
Further increasing the number of direct detections of high-$z$ DLA galaxies will enable us to determine the relationship between DLAs and other populations, star-formation processes in a galactic halo, and the feedback to the IGM.
The large variety of 2D spectral shape among the observed sample of \citet{rau08} and ours could be due to a viewing angle effect, as suggested by \citet{bar11}, which demonstrated similar variations based on their 3D radiative transfer simulation.  
Ly$\alpha$ photons tend to escape through low-density paths in the circumgalactic medium; therefore, Ly$\alpha$ morphology is strongly dependent on a clumpy distribution of H {\sc i} clouds.
Our interpretation of extended Ly$\alpha$ emission due to the outflow in a clumpy ISM is consistent with the picture implied by \citet{law12b} for star-forming galaxies and does not necessarily support a classical bipolar outflow, which is often probed by Mg {\sc ii} absorption at low-z \citep{bou12, lun12}.  
Spatially extended Ly$\alpha$ emission also strongly depends on the inflow/outflow implementation; conversely, it provides an important diagnostic to infer the relationship between the internal dynamics of high-z galaxies and the circumgalactic medium.
Although our sample is only one DLA galaxy, detection of certain emissions from these heavy absorbers would reveal the relation between  ^^ ^^ lights" and ^^ ^^ shades" in the high-$z$ universe.

\acknowledgments

We thank the referee for his/her helpful comments that improved the manuscript.
We thank Yuichi Matsuda for their useful discussions.
We are grateful to the Subaru Observatory staffs for their help with the observations.
This research was supported by the Japan Society for the Promotion of Science through Grant-in-Aid for Scientific Research 23340050.



{\it Facilities:} \facility{Subaru (FOCAS)}.

\end{document}